\documentclass{article}

\usepackage{graphicx}

\begin{document}
\date{}
\def\be{\begin{equation}}
\def\ee#1{\label{#1}\end{equation}}

\def\lle{\left[\hskip-1pt\left[}
\def\rre{\right]\hskip-1pt\right]}

\title{Brane cosmology with a van der Waals \\
equation of state}
\author{G. M. Kremer\thanks{E-mail: 
kremer@fisica.ufpr.br}\\Departamento de F\'\i sica, 
Universidade Federal do Paran\'a,\\
Caixa Postal 19044, 81531-990 Curitiba, Brazil}
\maketitle

\begin{abstract}
The evolution of  a Universe confined onto a 3-brane embedded in a 
five-dimensional space-time is investigated where the cosmological fluid 
on the brane is modeled by the van der Waals  equation of state. It is 
shown that the Universe on the brane 
evolves in such a manner that three distinct periods concerning its
acceleration field are attained: (a) an  initial accelerated epoch where the 
van der Waals fluid behaves like a scalar field with a negative pressure;
(b) a past decelerated period  which has two contributions, 
one of them is related to the
van der Waals fluid which behaves like a matter field with a positive pressure,
whereas the other contribution comes from a term of the 
Friedmann equation on the brane
which is inversely proportional to the scale factor to the fourth power
and can be interpreted as a radiation field, and  (c) a present 
accelerated phase due to a cosmological constant on the brane.

\end{abstract}


According to the cosmological observations one can distinguish 
three distinct periods
for the Universe that are related to its acceleration field. 
The first period refers to an accelerated epoch dominated by a scalar field
where a rapid expansion of the Universe characterizes its inflationary phase. 
The next one is related to a decelerated phase dominated by  matter fields.
This  period is followed by a return to an accelerated epoch
dominated by a cosmological constant or dark energy.

Recently several authors have investigated the evolution of the Universe
within the framework of the one-brane model of Randall and Sundrum~\cite{RS1}
where the Universe is confined onto a 3-brane, which is a hyper-surface 
embedded in a five-dimensional space-time called bulk. For recent reviews 
on this subject one is referred to the works of Brax and van de 
Bruck~\cite{BB} and Langlois~\cite{Lan} and the references therein.

In the present work we investigate the evolution of  a Universe confined 
onto the 3-brane embedded in a five-dimensional
space-time in order to describe the three
distinct periods of the Universe beginning with an accelerated phase passing 
through a decelerated epoch and returning to an accelerated period.
For that end we model the cosmological fluid on the brane by the van der Waals 
equation of state. The use
of the van der Waals equation of state in cosmological problems was  
first proposed
by Capozziello and co-workers~\cite{CMF,CCT}, who recognize that this equation
of state could describe the transition from a scalar field dominated period
to a matter dominated epoch without the need of introducing scalar fields. 
Recently a model for the Universe as a mixture of a van der Waals fluid 
with dark energy (modeled as quintessence or Chaplygin gas)
was proposed in the work~\cite{K2} -- within the framework of a 
four-dimensional space-time theory -- in order to describe the transition 
from the accelerated-decelerated-accelerated periods of the Universe.

In this work  we show that the Universe on the brane modeled by the van der 
Waals equation of state evolves in such a manner that the three distinct 
periods concerning its
acceleration field are attained. The initial accelerated epoch is due to the 
van der Waals fluid which behaves like a scalar field with a negative pressure.
The past decelerated period  has two contributions: one of them is 
related to the van der Waals fluid which behaves like a matter field 
with a positive pressure,
whereas the other comes from  a term of the Friedmann equation on the brane
which is inversely proportional to the scale factor to the fourth power
and can be interpreted as a radiation field. The present accelerated period
is due to a cosmological constant on the brane. 

In the one-brane model of Randall and Sundrum~\cite{RS1} the 
Universe is confined onto a hyper-surface -- called brane -- which is 
embedded in a five-dimensional space-time with coordinates
$(t,x^1,x^2,x^3,y)$, called bulk. The brane is located at $y=0$ 
and the line element which describes a spatially flat, homogeneous and 
isotropic Universe on the brane
is given by (see, for example, the reviews~\cite{BB,Lan})
\be
ds^2=g_{MN}dx^Mdx^N=n(t,y)^2dt^2-a(t,y)^2\delta_{ij}dx^idx^j-dy^2,
\ee{b1}
where $g_{MN}$ is the metric tensor with signature $(+,-,-,-,-)$. 
The two functions $n(t,y)$ and $a(t,y)$ are determined from the 
five-dimensional Einstein field equations  that read
\be
R_{MN}-{1\over 2}Rg_{MN}+\Lambda g_{MN}=-\kappa^2 T_{MN}.
\ee{b2}
Above, $R_{MN}$ is the five-dimensional Ricci tensor, $R={R^M}_M$ its trace,  
$\Lambda$ denotes a bulk cosmological constant, $\kappa^2=8\pi G_5$ is 
related to the five-dimensional gravitational constant $G_5$, and $T_{MN}$ is
the five-dimensional energy-momentum  tensor.

The five-dimensional energy-momentum  tensor $T_{MN}$  is decomposed 
into a sum of two terms: one refers
to the bulk whereas the other is related to the brane, i.e.,
\be
{T^M}_N={T^M}_N\vert_{\rm bulk}+{T^M}_N\vert_{\rm brane}.
\ee{b7}
The energy-momentum tensor in the bulk is given by
\be
{T^M}_N\vert_{\rm bulk}={\rm diag}\, (\rho_B, -p_B, -p_B, -p_B, \rho_B),
\ee{b8}
where the energy density $\rho_B$ and the pressure $p_B$ in the 
bulk do not depend on the $y$ coordinate. Furthermore, the energy-momentum 
tensor on the brane reads
\be
T_{MN}\vert_{\rm brane}=\delta(y)(\sigma g_{\mu\nu}+{\cal T}_{\mu\nu})
{\delta^\mu}_M {\delta^\nu}_N.
\ee{b9}
On the brane there exist two contributions: one is related to the 
constant tension on the brane $\sigma$, whereas the other refers to the
energy-momentum tensor of the cosmological fluid ${\cal T}_{\mu\nu}$ 
which is written as
\be
{{\cal T}^\mu}_\nu={\rm diag}\, (\rho, -p, -p, -p).
\ee{b10}
Above, $\rho$ and $p$ denote the energy density and the pressure of the 
cosmological fluid on the brane, respectively.

For the line element given by (\ref{b1}),  the components of the 
Einstein field equations (\ref{b2}) become
\be
3\left\{\left({\dot a\over a}\right)^2- n^2\left[{a''\over a}+
\left({a'\over a}\right)^2\right]\right\}-\Lambda n^2 =\kappa^2 T_{00},
\ee{b3}
$$
\left\{a^2\left[\left({a'\over a}\right)^2+2{a'n'\over an}
+2{a''\over a}+{n''\over n}\right]-{a^2\over n^2}
\left[\left({\dot a\over a}\right)^2
-2{\dot a\dot n\over an}+2{\ddot a\over a}\right]\right\}\delta_{ij}
$$
\be
+\Lambda a^2\delta_{ij}
=\kappa^2 T_{ij},
\ee{b4}
\be
3\left({\dot a\over a}{n'\over n}-{\dot a'\over a}\right)=\kappa^2 T_{0y},
\ee{b5}
\be
3\left({a'\over a}\right)^2+3{a'n'\over an}-
{3\over n^2}\left[\left({\dot a\over a}\right)^2-{\dot a\dot n\over an}
+{\ddot a\over a}\right]+\Lambda=\kappa^2 T_{yy}.
\ee{b6}
In the above equations the dot and the prime refer to a differentiation 
with respect to the coordinates $t$ and $y$, respectively.

We follow the work~\cite{BDL} and introduce the
abbreviations at $y=0$: $a_0(t)\equiv a(t,y=0)$ and 
$n_0(t)\equiv n(t,y=0)$. Hence, on the brane the line element is written as 
\be
ds^2=n_0(t)^2dt^2-a_0(t)^2\delta_{ij}dx^idx^j.
\ee{b11}
By imposing the gauge condition $n_0(t)=1$, the time coordinate 
becomes the proper time on the brane and  $a_0(t)$ is identified
with the scale factor.

If one assumes that there exists no matter flow along the fifth dimension  
(see, e.g.~\cite{BDEL})-- so that $T_{0y}=0$ -- 
 the integration of equation (\ref{b5}) leads to
\be
n(y,t)={\dot a(t,y)\over \dot a_0(t)}.
\ee{b12}

We integrate equations (\ref{b3}) and (\ref{b4}) over 
$-\epsilon\leq y\leq +\epsilon$, and by taking the limit 
$\epsilon\rightarrow0$, it follows the so-called junction conditions:
\be
{\lle a' \rre \over a_0}=-{\kappa^2\over 3}(\rho+\sigma),\qquad
\lle n' \rre =-{\kappa^2\over 3}\sigma+{2\kappa^2\over 3}\rho
+\kappa^2p,
\ee{b13}
where $\lle f\rre=f(0+)-f(0-)$ denotes the jump of the function 
$f$ across the brane.

If we take into account in  equation (\ref{b5}) the junction 
conditions (\ref{b13}) together with the condition of no matter 
flow along the fifth dimension -- i.e., $T_{0y}=0$ -- we get
the conservation equation for the energy density on the brane
\be
\dot \rho+3{\dot a_0\over a_0}(\rho+p)=0.
\ee{b14}

In order to analyze the system of Einstein field equations 
(\ref{b3}) -- (\ref{b6}) in the bulk, we follow the work~\cite{BDEL} 
and introduce the function
\be
F(t,y)\equiv (a'a)^2-{(\dot a a)^2\over n^2}.
\ee{b15}
In terms of the function $F(t,y)$ the 00, $yy$ and  $ij$ - components 
in the bulk read:
\be
F'=-{2a'a^3\over 3}(\Lambda+\kappa^2\rho_B),\qquad
\dot F=-{2\dot a a^3\over 3}(\Lambda+\kappa^2\rho_B),
\ee{b16}
\be
\left({F'\over a'}\right)^\bullet=-2\dot a a^2(\Lambda-\kappa^2p_B),
\ee{b17}
respectively. Now we differentiate (\ref{b16})$_1$ with respect to time $t$ 
and  (\ref{b16})$_2$ with respect to the coordinate $y$, and  get that 
$\rho_B$ does not depend on time $t$. Furthermore, from the differentiation
with respect to time
of the expression $F'/a'$ obtained from  (\ref{b16})$_1$ 
 it follows an equation  when compared with 
(\ref{b17}) implies that  $p_B=-\rho_B$. Finally, the integration of
(\ref{b16})$_1$ with respect to the coordinate $y$, leads to
\be
F=(a'a)^2-{(\dot a a)^2\over n^2}=-{a^4\over 6}(\Lambda+\kappa^2\rho_B) +C.
\ee{b18}
Above, $C$ is a constant since $\rho_B$ is time independent.

The Friedmann equation on the brane  is obtained from (\ref{b18}) 
by considering the limit $y\rightarrow0$ and the junction conditions 
(\ref{b13}), yielding
\be
\left({\dot a_0\over a_0}\right)^2={\kappa^4\over 18}\sigma\rho+
{\kappa^4\over 36}\rho^2+{1\over 36}(\kappa^4\sigma^2+6\Lambda+6\kappa^2\rho_B)
+{C\over a_0^4}.
\ee{b19}

From the Friedmann equation one can determine the acceleration equation on 
the brane. Indeed,  the differentiation of (\ref{b19}) with respect to time,
yields
\be
{\ddot a_0\over a_0}=-{\kappa^4\over 36}\sigma(\rho+3p)
-{\kappa^4\over 36}\rho(2\rho+3p)+{1\over 36}
(\kappa^4\sigma^2+6\Lambda+6\kappa^2\rho_B)-{C\over a_0^4},
\ee{b20}
thanks to the  equation (\ref{b14}) for the energy density on 
the brane. This equation can  be  obtained  also from 
equation (\ref{b6}) by using the junction conditions (\ref{b13}).

Now we have a system of differential equations for the determination 
of the energy density $\rho(t)$ and of the scale factor $a_0(t)$  on the
brane which is composed by the conservation equation for the
energy density (\ref{b14}) and by the Friedmann equation (\ref{b19})  -- 
or by the 
acceleration equation (\ref{b20}). 
In order to find the time evolution of the energy density  and of the
scale factor from the system of differential equations
one has to prescribe initial conditions for these fields and 
to close the system  by choosing an equation of state  which relates the
pressure of the cosmological fluid to the energy density on the brane, i.e,
$p=p(\rho)$. 

For the determination of the solution of the system of differential equations
described above we introduce the 
dimensionless quantities 
\be 
t\equiv {\kappa^2\sigma\over \sqrt{18}}t,\qquad
\rho\equiv{\rho\over\sigma},\qquad p\equiv{p\over\sigma},\qquad
a_0\equiv{a_0(t)\over a_0(0)},
\ee{b21}
and write the conservation equation for the
energy density (\ref{b14}), the
Friedmann (\ref{b19}) and the acceleration 
(\ref{b20}) equations in terms of the  dimensionless quantities as
\be
\dot \rho+3{\dot a_0\over a_0}(\rho+p)=0,
\ee{b14a}
\be
\left({\dot a_0\over a_0}\right)^2=\rho+
{\rho^2\over 2}+{\lambda\over3}+{\chi\over a_0^4},
\ee{b22}
\be
{\ddot a_0\over a_0}=-{1\over 2}(\rho+3p)
-{\rho\over2}(2\rho+3p)+{\lambda\over3}-{\chi\over a_0^4}.
\ee{b23}
In equations (\ref{b22}) and (\ref{b23}) the following abbreviations 
were introduced
\be
\lambda=9\left[{\Lambda\over \kappa^4\sigma^2}+{\rho_B\over\kappa^2\sigma^2}
+{1\over6}\right],\qquad
\chi={18C\over [a_0(0)\kappa]^4\sigma^2}.
\ee{b24}
The constant $\lambda$ in the Friedmann (\ref{b22}) and acceleration 
(\ref{b23}) 
equations can be interpreted as a cosmological constant on the brane. Moreover,
it is noteworthy to call attention to the fact that equations 
(\ref{b22}) and (\ref{b23}) can be written as
\be
\left({\dot a_0\over a_0}\right)^2=\rho+\rho_{\rm rad}+
{\rho^2\over 2}+{\lambda\over3},
\ee{b25}
\be
{\ddot a_0\over a_0}=-{1\over 2}(\rho+\rho_{\rm rad}
+3p+3p_{\rm rad})
-{\rho\over2}(2\rho+3p)+{\lambda\over3},
\ee{b26}
thanks to the well-known relationships valid in   
the four-dimensional case where
the energy density of the radiation field $\rho_{\rm rad}$ scales as 
$1/a_0^4$ and the radiation pressure is 
related to the energy density by $p_{\rm rad}=\rho_{\rm rad}/3$. 

From now on we shall analyze the system of differential equations
consisting of the  conservation equation for the
energy density (\ref{b14a}) and of the acceleration equation (\ref{b23}).      
In order to close this system of differential equations 
we have to choose an equation of state that relates the pressure
to the energy density of the cosmological fluid on the brane. In the 
four-dimensional case one normally uses a barotropic equation of 
state $p=w\rho$ with $0\leq w\leq1$ to represent a radiation or matter 
dominated Universe. The barotropic equation of state with $-1\leq w\leq0$ 
is also used to represent the scalar fields: the inflaton in the inflationary 
period of the Universe, and the quintessence in the dark energy dominated 
epoch of the Universe. Other equations of state are also used to model the 
cosmological fluid, namely the Chaplygin equation of state (see e.g. the 
work~\cite{KC} and the references therein) which  can describe a 
transition from a matter dominated period to a cosmological constant 
dominated epoch and the  van der Waals equation of state which can simulate
the transition from an inflationary period to a matter field dominated  
epoch~\cite{K2}. Here we interested in the description of an  
accelerated  period followed
by a decelerated epoch, hence we shall use the van  der Waals equation of state
\be
p={8w\rho\over 3-\rho}-3\rho^2,
\ee{b27}
where the parameter $w$ could be identified with the coefficient of 
proportionality in the barotropic formula, since for small values of 
the energy density $p\propto w\rho$.  In classical thermodynamics 
(see e.g.~\cite{Callen})
the equation (\ref{b27}) is a reduced van der Waals equation of state 
where the free parameter $w$ is connected with a dimensionless temperature.
Capozziello and co-workers~\cite{CMF,CCT} have used a  non-reduced form 
of the van der Waals
equation which is characterized by three free parameters instead of only one. 
In their works~\cite{CMF,CCT} they also write a reduced form of the van der 
Waals equation  with only one free parameter which is not similar to the 
reduced van der Waals equation  of state normally found in the literature, 
since the free parameter multiplies also the  term proportional to $\rho^2$. 

Apart from the initial conditions for the fields of energy 
density $\rho(0)$, scale factor $a_0(0)$ and velocity $\dot a_0(0)$, 
one has to specify values for the parameters $\lambda$, $\chi$ and 
$w$ in order to find a solution of the system of differential 
equations (\ref{b14a}) and (\ref{b23}) which is closed by the van der Waals 
equation of state (\ref{b27}). Here we have specified the 
initial conditions (by adjusting clocks): $\rho(0)=1$ for the energy density, 
$a_0(0)=1$ for the  scale factor and
$\dot a(0)=\sqrt{3/2+\lambda/3+\chi}$ for the velocity field, which is a 
consequence of the Friedmann equation (\ref{b22}). Furthermore,  in order 
to plot the time evolution of the fields in figures 1 and 2
we have chosen the following values for the parameters
$\lambda=0.03$, $\chi=0.5$, and three different values for 
$w$, namely, $w=0.51$, $w=0.501$, and $w= 0.5001$.  
Later on we shall discuss how the changes in 
the parameters $w$, $\lambda$ and $\chi$ affect the solutions of the system 
of differential equations.

\begin{figure}\begin{center}
\includegraphics[width=8.5cm]{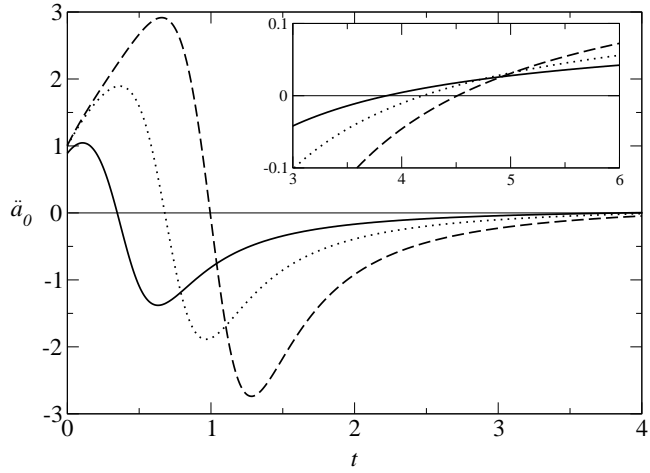}
\caption{Acceleration  $\ddot a_0$
vs time $t$ for $w=0.51$ (straight line), for $w=0.501$ (dotted line), 
and for $w=0.5001$ (dashed line).}
\end{center}\end{figure}
 
\begin{figure}\begin{center}
\includegraphics[width=8.5cm]{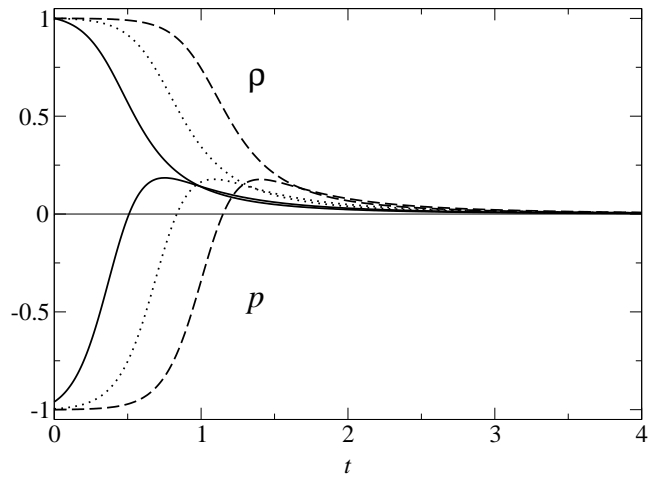}
\caption{Energy density $\rho$ 
and pressure $p$ 
vs time $t$ for $w=0.51$ (straight lines), for $w=0.501$ (dotted lines), 
and for $w=0.5001$ (dashed lines).}
\end{center}\end{figure}

In figure 1 it is plotted the acceleration field  $\ddot a_0(t)$
as function of time $t$ for $w=0.51$ (straight line),
$w=0.501$ (dotted line), and $w=0.5001$ (dashed line). 
We infer from this figure that the acceleration field for these three values of
$w$ describes the distinct accelerated-decelerated-accelerated phases of 
the Universe. 
In the first period, which refers to a scalar field dominated 
epoch, the positive acceleration grows up to a maximum value followed 
by a decay towards zero. The next phase is related to a matter dominated 
period where the acceleration is always negative and decays to a maximum 
negative value followed by a growth towards zero. The third period is a 
cosmological constant dominated epoch where 
the acceleration field assumes a positive value. 
The  energy density $\rho$ and pressure $p$  fields 
are plotted in figure 2 as function of time $t$ for $w=0.51$ (straight line),
$w=0.501$ (dotted line), and $w=0.5001$ (dashed line). We 
conclude from this figure that for the three values of $w$ 
the energy density field
decays with time whereas the  pressure field has two distinct behaviors. 
At the beginning the pressure is negative so that  the van der 
Waals fluid behaves like a scalar field and it is the responsible for the 
initial accelerated period. At later times the pressure becomes positive 
so that the van der Waals fluid behaves like a matter field and it is partially
responsible for the decelerated period, since the radiation field which scales 
as $1/a_0^4$  also contributes to the decelerated phase of the Universe. 
When  $w$ approaches the value of $w=0.5$, we can also infer from the 
figures  that the pressure field of
the van der Waals fluid at the beginning  behaves like an inflaton with an 
equation of state given by $p=-\rho$ and it is the responsible for an increase 
of the early acceleration.  

We have chosen three nearby values for $w$ in order to show how 
the acceleration, energy density and pressure fields behave for small 
changes of $w$. We proceed now to discuss how significant changes in the values
of $w$ have influence on the behavior of these fields for  fixed values
of $\lambda$ and $\chi$, here $\lambda=0.03$ and $\chi=0.5$.
For $w=0.5$ the energy density remains constant
so that the van der Waals fluid  behaves like an inflaton with an
equation of state given by $p=-\rho$,  the 
acceleration field grows exponentially and in this case there exists only an 
accelerated phase for the Universe. For values of $w<0.5$ the energy 
density grows with time, and we infer that this 
behavior does not represent a physical solution for the 
evolution of the Universe.  
For values of $0.5<w<0.58$ the initial 
accelerated period decreases, since the negative part
of pressure of the van der Waals fluid 
decreases, whereas its positive part increases. Hence in the interval 
$0.5<w<0.58$ the three periods accelerated-decelerated-accelerated are 
present in the evolution of the Universe. These three periods for the 
Universe can be found for a larger interval for $w$ by decreasing the 
amount of radiation, i.e., by decreasing 
the value of the constant $\chi$. For values of $w>0.58$  (and $\lambda=0.03$,
$\chi=0.5$) the    acceleration field evolves from a matter dominated 
Universe where $\ddot a<0$ to a cosmological constant dominated Universe 
where $\ddot a>0$. In this last case there exists no inflationary period.

Let us comment on the behavior of the fields when we change the values of 
the parameters $\chi$ and $\lambda$. By increasing the value of $\chi$ 
there exists a more pronounced predominance of the  radiation field 
which scales as $1/a_0^4$ and the deceleration period begins at earlier times.
By decreasing the value of $\lambda$, which is connected to the cosmological 
constant on the brane, the present accelerated period begins at later times.
Moreover, the limit $ \lambda\rightarrow0$ leads to  an initial accelerated 
period followed by a  decelerated period without a present accelerated period,
indicating that $\lambda$ plays the role of the dark energy. We could also 
obtain the present accelerated period -- which is characterized by a 
dark energy  dominated Universe -- by following the same methodology 
of the work~\cite{K2} and  instead of  introducing a cosmological constant 
which represents the dark energy, we could model the dark energy 
by an equation of state like the Chaplygin or the quintessence
equations of state.

As a final comment we have  also investigated the solution of the system
of differential equations (\ref{b14a}) and (\ref{b23}) closed  by a barotropic 
equation of state for the cosmological fluid 
on the brane, i.e., $p=w\rho$ with $-1\leq w\leq1$. 
The results we have found for the acceleration field by changing
the values of $w$ describe either an accelerated phase, or a 
decelerated epoch followed by an accelerated period of the Universe, 
i.e., with the
barotropic equation of state it was not possible to describe the three phases
accelerated-decelerated-accelerated found here by using the 
van der Waals equation of state.

\end{document}